\documentclass[%
 aip,
 jmp,%
 amsmath,amssymb,
preprint,%
]{revtex4-1}

\usepackage{graphicx}
\usepackage{dcolumn}
\usepackage{bm}
\usepackage{color}

\begin{document}


\title{Collisionless electrostatic shock formation and ion acceleration in intense laser interactions with near critical density plasmas}
\author{M. Liu}
\affiliation{Key Laboratory for Laser Plasmas (MoE), Department of Physics and Astronomy, Shanghai Jiao Tong University, Shanghai 200240, China}%
\affiliation{Collaborative Innovation Center of IFSA, Shanghai Jiao Tong University, Shanghai 200240, China}%
\author{S. M. Weng}
 \email{wengsuming@gmail.com}%
\affiliation{Key Laboratory for Laser Plasmas (MoE), Department of Physics and Astronomy, Shanghai Jiao Tong University, Shanghai 200240, China}%
\affiliation{Collaborative Innovation Center of IFSA, Shanghai Jiao Tong University, Shanghai 200240, China}%
\author{Y. T. Li}
\affiliation{Collaborative Innovation Center of IFSA, Shanghai Jiao Tong University, Shanghai 200240, China}%
\affiliation{National Laboratory for Condensed Matter Physics, Institute of Physics, Chinese Academy of Sciences, Beijing 100190, China}%
\author{D. W. Yuan}
\affiliation{Key Laboratory of Optical Astronomy, National Astronomical Observatories, CAS, Beijing 100012, China}%
\author{M. Chen}
\affiliation{Key Laboratory for Laser Plasmas (MoE), Department of Physics and Astronomy, Shanghai Jiao Tong University, Shanghai 200240, China}%
\affiliation{Collaborative Innovation Center of IFSA, Shanghai Jiao Tong University, Shanghai 200240, China}%
\author{P. Mulser}
\affiliation{Theoretical Quantum Electronics (TQE), Technische Universit\"{a}t Darmstadt, D-64289 Darmstadt, Germany}%
\author{Z. M. Sheng}
 \email{zmsheng@sjtu.edu.cn}%
\affiliation{Key Laboratory for Laser Plasmas (MoE), Department of Physics and Astronomy, Shanghai Jiao Tong University, Shanghai 200240, China}%
\affiliation{Collaborative Innovation Center of IFSA, Shanghai Jiao Tong University, Shanghai 200240, China}%
\affiliation{SUPA, Department of Physics, University of Strathclyde, Glasgow G4 0NG, UK}
\author{M. Murakami}
\affiliation{Institute of Laser Engineering, Osaka University, Osaka 565-0871, Japan}%
\author{L. L. Yu}
\affiliation{Key Laboratory for Laser Plasmas (MoE), Department of Physics and Astronomy, Shanghai Jiao Tong University, Shanghai 200240, China}%
\affiliation{Collaborative Innovation Center of IFSA, Shanghai Jiao Tong University, Shanghai 200240, China}%
\author{X. L. Zheng}
\affiliation{Key Laboratory for Laser Plasmas (MoE), Department of Physics and Astronomy, Shanghai Jiao Tong University, Shanghai 200240, China}%
\affiliation{Collaborative Innovation Center of IFSA, Shanghai Jiao Tong University, Shanghai 200240, China}%
\author{J. Zhang}
\affiliation{Key Laboratory for Laser Plasmas (MoE), Department of Physics and Astronomy, Shanghai Jiao Tong University, Shanghai 200240, China}%
\affiliation{Collaborative Innovation Center of IFSA, Shanghai Jiao Tong University, Shanghai 200240, China}%

\date{\today}
\begin{abstract}
Laser-driven collisonless electrostatic shock formation and the subsequent ion acceleration have been studied in near critical density plasmas. Particle-in-cell simulations show that both the speed of laser-driven collisionless electrostatic shock and the energies of shock-accelerated ions can be greatly enhanced due to fast laser propagation in near critical density plasmas. However, a response time longer than tens of laser wave cycles is required before the shock formation in a near critical density plasma, in contrast to the quick shock formation in a highly overdense target. More important, we find that some ions can be reflected by the collisionless shock even if the electrostatic potential jump across the shock is smaller than the ion kinetic energy in the shock frame, which seems against the conventional ion-reflection condition. These anomalous ion reflections are attributed to the strongly time-oscillating electric field accompanying laser-driven collisionless shock in a near critical density plasma.
\end{abstract}

\pacs{52.38.Kd, 41.75.Jv, 52.50.Jm, 52.65.Rr}
\maketitle

\section{\label{sec:level1}Introduction}

Collisionless shocks driven by intense laser pulses attract growing attention from both laboratory astrophysics \cite{Kuramitsu2012, Stockem2014,Bret} and plasma-based accelerator communities \cite{Haberberger}. In astrophysics, collisionless shocks are observed in supernova remnants, gamma ray bursts, active galactic nucleus jets and so on \cite{Bamba}. Particularly, collisionless shocks are relevant to the generation of extremely high energy particles, which is one of the most fundamental problems in astrophysics \cite{Spitkovsky, Martins}. However, the direct study of collisionless shocks in astronomical environments is passive and suffers from the lack of details. In contrast, high power laser-matter interaction can create astrophysical-like conditions and thus provides convenient test-beds for many astrophysical processes including the collisionless shock formation \cite{Kuramitsu2012,Stockem2014,Bret,Xun,Yuan,Ahmed}. With regard to plasma-based accelerator, ion accelerations by laser-driven collisionless electrostatic shocks have been observed in a great number of simulations \cite{Denavit, Silva, Macchi, He, Chen, Chen2} and experiments \cite{Wei,Zhang}. In particular, monoenergetic ion acceleration by collisionless shock has been proposed by using a near critical density target with a slowly decreasing density profile at the rear side \cite{Fiuza2012,Fiuza2013}. With this target configuration, the electric field relevant to the target normal sheath acceleration becomes smaller and nearly constant. Recently, quasi-monoenergetic ion beams have been obtained in the experiments using infrared CO2 laser pulses and gas-jet targets \cite{Palmer, Haberberger, Tresca}. The gas-jet targets have the advantage of operating at a high-repetition rate, which is a crucial requirement in many practical applications such as cancer radiotherapy \cite{Zeil}.

In principle, collisionless electrostatic shocks can originate from the temperature or/and density jumps of plasmas \cite{Forslund}. The kinetic theory of collisionless electrostatic shock formation in the interaction of plasma slabs with different temperatures and densities was presented by Sorasio \emph{et al.}\cite{Sorasio}, and its relativistic generalization and ion-reflection condition were given by Stockem \emph{et al.}\cite{Stockem2013}.
For a rigorous electrostatic shock, the ion-reflection condition can be written as \cite{He,Stockem2013}
\begin{equation}
Z_i e \Phi_{\max} \geq m_iv_{i}^2/2, \label{ref-condition}
\end{equation}
where $Z_i$ is the ionic charge state, $\Phi_{\max}$ is the peak of electrostatic potential, $m_iv_{i}^2/2$ is the ion kinetic energy in the shock frame, and for initially cold ions $|v_{i}|$ are equal to the shock speed $v_{shock}$.
In the interaction of an intense laser pulse with a highly overdense target, the shock speed is mainly determined by the laser hole-boring (or piston) velocity $v_{HB}\simeq \sqrt{I/m_in_ic^3}$ \cite{Silva,Chen}.
While laser-driven collisionless shock has been extensively studied using highly overdense targets \cite{Denavit, Silva, Macchi, He, Chen, Zhang, Fiuza2012,Fiuza2013},
the laser-plasma interactions are more complicated in the near critical density regime \cite{WengNJP,Dover} and the collisionless shocks excited therein are rather unexplored \cite{Kim}.

In this paper, we study the formation of collisionless electrostatic shock and the subsequent ion acceleration in the interaction of intense laser pulses with near critical density plasmas.
A substantial number of electrons in such a plasma can be pushed forward due to the laser pondermotive force and form an electron density peak ahead of the laser front.
Simultaneously, a strong electrostatic field is generated due to the charge separation.
Consequently, some ions may be pulled by this charge separation field and be accumulated ahead of the laser front.
Meanwhile, the plasma will be greatly heated because of its strong coupling with laser.
Therefore, a high-temperature high-density plasma flow can be produced ahead of the laser front.
When such a laser-driven plasma flow interacts with the background plasma, a collisionless electrostatic shock can be excited. Particle-in-Cell simulations show that the speed of this collisionless shock can be very high since the laser pulse propagates much faster in a near critical density plasma than in a highly overdense plasma.
Most important, unexpected ion reflections by this collisionless shock is clearly observed even if the conventional ion-reflection condition Eq.(\ref{ref-condition}) are not satisfied.

\section{Simulation results}

To study the laser-driven collisionless shock generation in the near critical density regime, 1D3V particle-in-cell (PIC) simulations have been performed with the code OSIRIS \cite{Fonseca}. The simulation box locates in $-20 \leq x/\lambda \leq 220$ and consists of 48,000 cells, and 600 macroparticles per cell are allocated in the plasma region, where $\lambda$ is the laser wavelength. In the simulations, initially cold and uniform targets are positioned in $0 \leq x/\lambda \leq 200$. The targets are assumed to be fully ionized, and the ions are protons.
For convenience, we assume that the laser pulse is linearly polarized (LP) in the $y$ direction and propagates along the positive $x$ direction.
For reference, the laser pulse is assumed to arrive at the target surface $x=0$ at $t=0$, and all simulations begin at $t=-20T_0$ and end at $t=200T_0$, where $T_0=\lambda/c$ is the wave period of the laser.
The intensity of the laser pulse increases to a constant value $a_0$ after a quick rising in the first five wave cycles, where $a_0\equiv |e\textbf{E}/m_e\omega c|$ is the normalized amplitude of laser electric field.
The target electron density $n_e$ varies from $n_c$ to $100n_c$ in different simulation cases, where $n_c \equiv m_e\omega^2/4\pi e^2$ is the critical density and $\omega=2\pi c/\lambda$ is the angular frequency of the laser.
We have found that the dynamics of laser-driven collisionless shock is very distinct in different laser-plasma parameter regimes.

\subsection{Shock formation in near critical density plasmas}

\begin{figure}
\centering
 \includegraphics[width=0.5\textwidth]{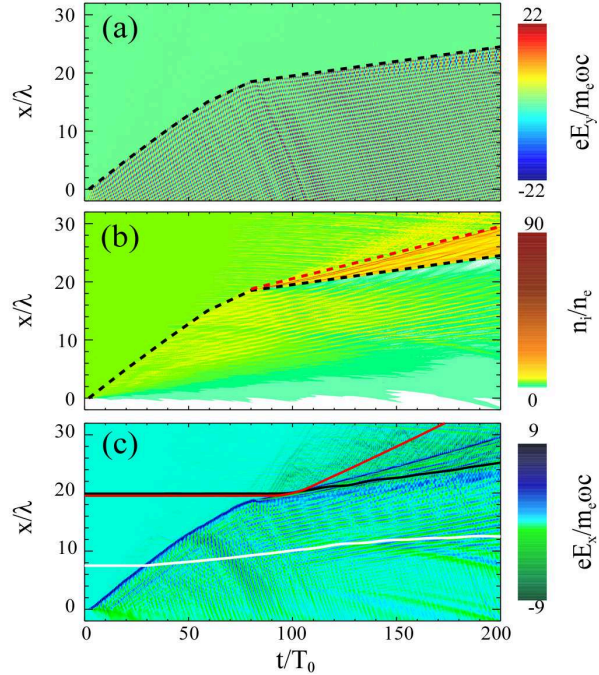}
\caption{Time evolutions of (a) transverse electric field $E_y$, (b) ion density $n_i$, and (c) longitudinal electric field $E_x$ in the interaction of an intense laser pulse of $a_0=10$ with a near critical density target of $n_e=5n_c$.
The black dashed lines in (a) and (b) mark the laser front, while the red dashed line and white dashed line in (b) marks the shock front.
The white, black, and red lines in (c) indicate the orbits of remaining, trapped, and shock-reflected ions, respectively.
} \label{figRT2D}
\end{figure}

We now analyze a representative simulation with an intense laser intensity $a_0=10$ and a near critical target density $n_e=5n_c$.
The relativistic-induced penetration of the laser pulse into this near critical density target is evidenced by the combination of Figs. \ref{figRT2D}(a) and \ref{figRT2D}(b), where the time evolutions of laser electric field $E_y$ and ion density $n_i$ are displayed, respectively.
The laser front is marked by the black dashed-line in Fig. \ref{figRT2D}(a), from which we can divide the laser propagation roughly into two stages.
In the first stage ($t \le 80T_0$), the forward velocity of the laser front $v_{laser}$ decreases gradually from $0.27c$ to $0.05c$.
In the second stage ($t \ge 80T_0$), the forward velocity of the laser front becomes nearly constant $v_{laser}\simeq 0.05c$. Moreover, the pronounced trapping and reflections of ions occur in the second stage as shown in Fig. \ref{figRT2D}(b).
\begin{figure}
\centering
 \includegraphics[width=0.5\textwidth]{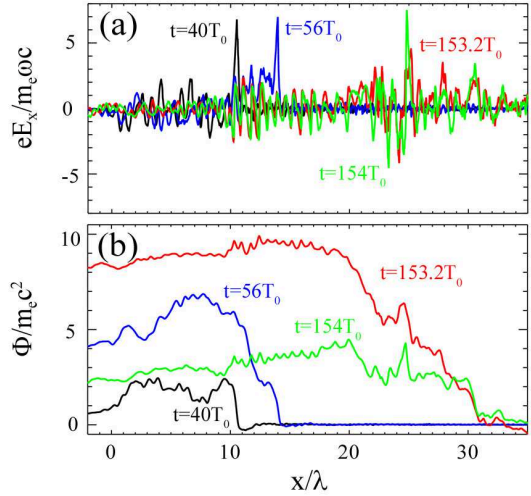}
\caption{(a) Longitudinal electric field $E_x$ and (b) electrostatic potential $\Phi$ at some different moments of $t=40$, 56, 153.2, and 154$T_0$ in the interaction of an intense laser pulse of $a_0=10$ with a near critical density target of $n_e=5n_c$.} \label{figExPhi}
\end{figure}

\begin{figure}
\centering
 \includegraphics[width=0.5\textwidth]{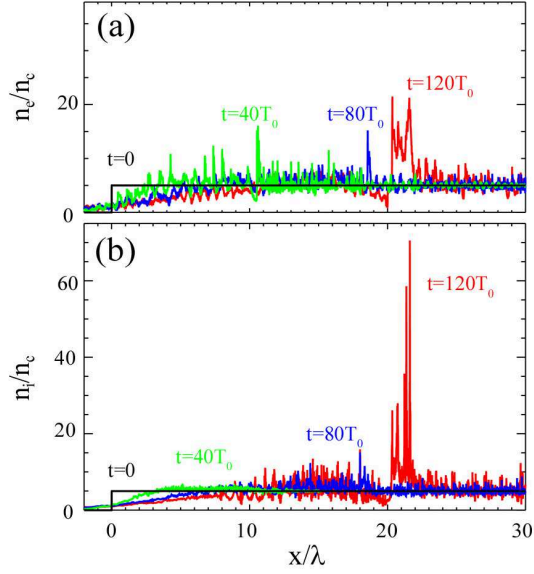}
\caption{(a) Electron density $n_e$ and (b) ion density $n_i$ at different moments of $t=0$, 40, 80, and 120$T_0$ in the interaction of an intense laser pulse of $a_0=10$ with a near critical density target of $n_e=5n_c$.} \label{figNeNi}
\end{figure}

To understand the distinct plasma dynamics in these two stages, we display the time evolution of longitudinal electric field $E_x$ in Fig. \ref{figRT2D}(c).
While the profiles of $E_x$ and the corresponding electrostatic potential $\Phi=\int_{+\infty}^{x} E_x dx$ at some representative moments are shown in Fig. \ref{figExPhi}(a) and \ref{figExPhi}(b), respectively.
In addition, the electron and ion density profiles at some representative moments are shown in Fig. \ref{figNeNi}(a) and Fig.\ref{figNeNi}(b), respectively.

Since the initial laser-plasma conditions satisfy $a=10>n_e/n_c=5$,
the relativistic transparency(RT) rather than the hole-boring(HB) dominates the laser propagation at the beginning of the first stage \cite{WengPoP}.
As a result, the initial forward velocity of the laser front $v_{laser} \simeq 0.27c$ is much higher than the hole-boring velocity $v_{HB}\simeq \sqrt{I/m_in_ic^3}\simeq 0.07c$.
Meantime, the laser pondermotive force will push a substantial number of electrons forward as shown in Fig. \ref{figNeNi}(a), resulting in a strong charge-separation field around the laser front.
Fig. \ref{figExPhi}(a) shows that the amplitude of this charge-separation field can be as high as $6.8 m_e\omega c/e$ at a representative moment $t=40T_0$. Despite of this strong electric field, no obvious ion reflection occurs at this moment as shown Fig. \ref{figNeNi}(b), and the white line in Fig. \ref{figRT2D}(c) shows a typical ion orbit.
This is because at this moment the laser front propagates at a speed as fast as $0.27c$, so does the charge-separation field.
As a result, the ion kinetic energy in the co-moving frame of the charge-separation field is about $m_iv_{laser}^2/2 \simeq 80 m_ec^2$, which is much larger than the peak of electrostatic potential $\Phi_{\max} \simeq 2.5 m_ec^2$ at this moment as shown in Fig. \ref{figExPhi}(b).
So no obvious ion reflection occurs at this moment.

However, the forward velocity of the laser front will gradually decrease due to the accumulation of particles ahead of the laser front and the dissipation of laser pulse in a dense plasma \cite{WengNJP,Robinson}.
As the ponderomotive force of a linearly polarized pulse is time oscillating, it cannot compress a tenuous plasma immediately if $a>n_e/n_c$.
However, some electrons will accumulate continually ahead of the laser front and form an electron density peak, thus a strong unipolar electrostatic field is formed.
After that some ions can accumulate ahead of the laser front too and form an ion density peak due to this strong unipolar electrostatic field.
Because the thermal velocity of the electrons are much faster than that of the ions, the ion density peak can be much narrower and higher than the electron density peak.
So the morphology of the charge separation field changes from unipolar to bipolar in this process.
It results in the further accumulation of the ions ahead of the laser front.
On the other hand, the accumulation of the particles also results in the broadening of the charge-separation layer, as indicated by the blue line in Fig.\ref{figExPhi}.
Due to the broadening of the charge-separation layer, the peak of the electrostatic potential $\Phi_{\max}$ increases to about $7 m_ec^2$ at this moment even if the amplitude of the charge-separation field is nearly same to the previous one.
For $\Phi_{\max} \simeq 7 m_ec^2$, we can estimate that it will be larger than the ion kinetic energy $m_iv_{laser}^2/2$ in the co-moving frame of the laser front if $v_{laser} < 0.087c$. Because the forward velocity of laser front $v_{laser}$ will decrease to about $0.05c$ at the end of this stage, the obvious trapping and reflections of ions will begin to occur at a turning point in this stage.
In return, the trapping and reflections of ions will dissipate the laser pulse and slow it down further.

In the second stage, however, it seems that a balance is roughly built between the propagation and the dissipation of the laser pulse as a nearly constant forward velocity of the laser front $v_{laser}\simeq 0.05c$ is presented.

Since the hole-boring rather than the relativistic transparency dominates the laser propagation in this stage \cite{WengNJP}, the forward velocity of the laser front is roughly equal to $ v_{HB} \simeq a_{\texttt{local}} \sqrt{m_e n_c/2 m_i n_e} c \simeq 0.05c$. Here, the amplitude of the local laser field $a_{\texttt{local}}\simeq 7$ is smaller than the amplitude of the laser field in the vacuum due to the dissipation in the plasma.

More importantly, we find that in the second stage the ion reflections may take place in two different scenarios. In the first scenario, a great number of ions can be reflected at the laser-plasma interface as in the first stage.  These reflected ions will form an ion beam that propagates faster than the laser pulse. During the propagation of this ion beam in the background plasma, a collisionless electrostatic shock can be stimulated with a strong longitudinal electric field at the beam-plasma interface as shown in Fig.\ref{figRT2D}(c).
At a representative moment $t=153.2T_0$, the peak of the electrostatic potential is about $10m_ec^2$ as shown in Fig. \ref{figExPhi}(b). This potential is even larger than the kinetic energy of background ions in the co-moving frame of the beam-plasma interface.
Therefore, in the second scenario, some ions in the background plasma can be reflected by the collisionless electrostatic shock that excited by the ion beam generated in the first scenario.
It is worthwhile to point out that the mean velocity of the ions reflected in the second scenario  ($\simeq 0.18c$) is about two times of that in the first scenario ($\simeq 0.09c$).
The mean velocity of the ions reflected in the first scenario is roughly twice the forward velocity of the laser front ($\simeq 0.05c$).
While the mean velocity of the ions reflected in the second scenario is roughly twice the velocity of the collisionless electrostatic shock.
As indicated by the slop of red dashed-line in Fig.\ref{figRT2D}(b), this shock velocity is about $0.09c$, which is roughly equal to the propagation velocity of the ion beam that consists of the ions reflected in the first scenario.


\subsection{Anomalous ion reflections by time-oscillating shock}

\begin{figure}
\centering
 \includegraphics[width=0.5\textwidth]{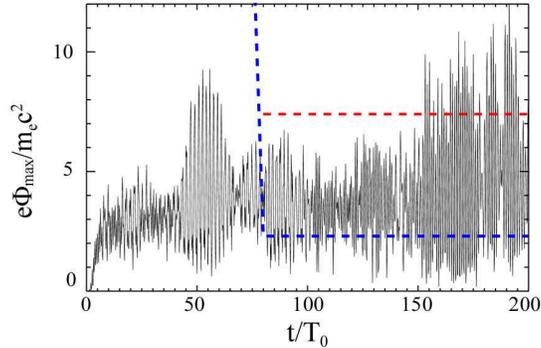}
\caption{Time evolutions of the peak of the electrostatic potential $\Phi_{\max}$ in the interaction of an intense laser pulse of $a_0=10$ with a near critical density target of $n_e=5n_c$.} \label{figphimax2D}
\end{figure}

Besides the ions being accelerated to higher energies, we notice that the longitudinal electric field is strongly time-oscillating in the interaction of an intense laser pulses with a near critical density plasma.
Although the time oscillation in the amplitude of the longitudinal electric field has been reported previously in the solitary or collisionless shock acceleration \cite{Macchipre}, here we find that not only the amplitude but also the profile shape of the longitudinal electric field oscillate violently is a near critical density plasma.
Fig. \ref{figExPhi}(a) shows that the field amplitude nearly doubles from $t=153.2T_0$ to $154T_0$, while the location of the field peak even steps backward. Consequently, the peak of the electrostatic potential also oscillates violently. As shown in Fig. \ref{figExPhi}(b), its value at $t=154T_0$ is even less than half of that at $t=153.2T_0$. This high-frequent nonlinear oscillation in the peak of the electrostatic potential is fully evidenced by Fig. \ref{figphimax2D}, which also illuminates that the conventional ion-reflection condition $\Phi \geq m_iv_{shock}^2/2$ can not always be satisfied after the shock formation ($t \ge 80T_0$).

\begin{figure}
\centering
 \includegraphics[width=0.5\textwidth]{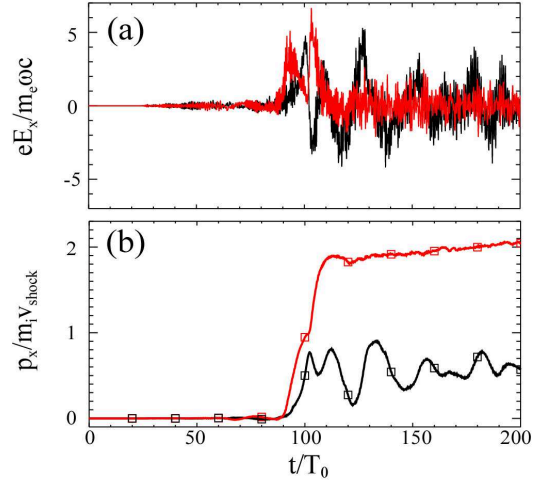}
\caption{Time evolutions of (a) the localized longitudinal electric fields $E_x[x_n(t),t]$ and (b)  the time integrals of these fields $e \int E_x[x_n(t),t] dt$ (i.e., the longitudinal momenta $p_x$) of the shock-reflected (red lines) and shock-trapped (black lines) ions, respectively, in the interaction of an intense laser pulse of $a_0=10$ with a near critical density target of $n_e=5n_c$.
The ion momenta obtained from the simulation are shown as the separate squares at some representative moments.
The orbits of the reflected and trapped ions are drawn in Fig.\ref{figRT2D}(c) as the red and black lines, respectively.
} \label{figExPx}
\end{figure}

However, we notice that the ion reflections by this collisionless shock are still possible even if the peak of the electrostatic potential $\Phi_{\max}$ is smaller than the ion kinetic energy $m_iv_{shock}^2/2$ in the shock frame. The underlying physics is that the longitudinal electric field here is a time-oscillating electrostatic field. Therefore, it does not need to satisfy the conventional ion-reflection condition in Eulerian coordinates. The general ion-reflection condition in a time-oscillating electric field can be obtained from the momentum conservation in Lagrangian coordinates as
\begin{equation}
p_x = e \int E_x[x_n(t),t] dt \geq m_i v_{shock}, \label{new-condition}
\end{equation}
 where $p_x$ is the longitudinal ion momentum, and $x_n(t)$ is the coordinate of the $n$-th ion at time $t$.

From the above inequality, we can see that the instantaneous electric field felt by the ion determines whether or not this ion is reflected. The red line in Fig. \ref{figRT2D}(c) illustrates an ion reflection by the shock at around $t\simeq 100 T_0$, despite that the peak of the electrostatic potential $\Phi_{\max}$ is much smaller than the ion kinetic energy $m_iv_{shock}^2/2$ during its reflection in Fig. \ref{figphimax2D}. Meanwhile, a lot of ions in the upstream region can not be reflected at the shock front, but they penetrate the shock and are trapped between the shock and the laser front.
The black line in Fig. \ref{figRT2D}(c) shows the orbit of such a trapped ion, which lag behind the shock front at latter time.
The accumulation of these trapped ions can result in the broadening of the charge-separation layer, which may contribute to the increase of the average value of the electrostatic potential peak $\Phi_{\max}$ at around $t=150T_0$.

The time evolutions of the localized longitudinal electric fields felt by the above mentioned shock-reflected and trapped ions are compared in Fig. \ref{figExPx}(a). The integrations of these two fields over the time result in the momenta of these two ions, which are displayed in Fig. \ref{figExPx}(b). As shown in Fig. \ref{figExPx}(a), the trapped ion experiences an alternatively positive and negative longitudinal electric field. Therefore, each of its acceleration periods is followed by a deceleration period to some extent in Fig. \ref{figExPx}(b). As a result, it can not be accelerated to very high energy and finally be trapped in the downstream. While the shock-reflected ion nearly always experiences a positive longitudinal electric field in Fig. \ref{figExPx}(a), hence it can be greatly accelerated. As shown in Fig. \ref{figExPx}(b), for the reflected ion the integration of the electric filed results in a momentum larger than $m_i v_{shock}$, and vice versa for the trapped ion.

\subsection{Effect of laser intensity}

\begin{figure}
\centering
 \includegraphics[width=0.5\textwidth]{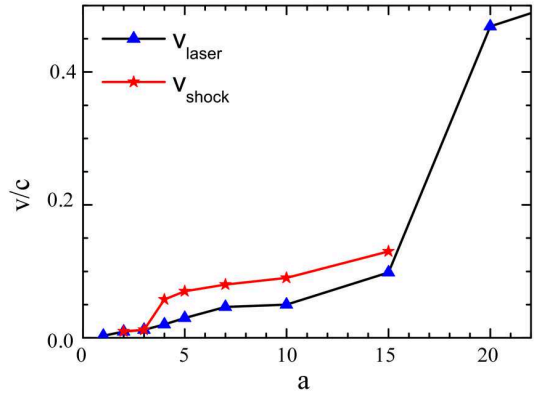}
\caption{The forward velocities of the laser front $v_{laser}$ (blue triangle-line) and the shock front  $v_{shock}$ (red star-line) as functions of the normalized amplitude of the incident laser intensity. The target parameters are same to those in Fig. \ref{figRT2D}.} \label{figvelocity}
\end{figure}

Due to the strong dissipation of the laser pulse in a near critical density target, the real-time local laser intensity when the collisionless shock is excited will be smaller than the initial incident laser intensity. Therefore, it may be interesting to investigate the effect of the laser intensity on the collisionless shock generation in a near critical density plasma. In particular, one may be curious about what will happen if we initially employ a laser pulse at the real-time local laser intensity.
For instance, the real-time local laser intensity is about $a_{local}\simeq 7$ when the collisionless shock is excited in the interaction of an intense laser pulse of $a_0=10$ with a near critical density target of $n_e=5n_c$. Using the initial laser intensity $a\simeq 7$, we find no qualitative difference in the shock generation except for a slightly lower shock velocity.
In Fig. \ref{figvelocity}, the forward velocities of the laser front $v_{laser}$ (blue triangle-line) and the shock front $v_{shock}$ (red star-line) are shown as functions of the normalized amplitude of the incident laser intensity.
One can see that the forward velocity of the laser front increases monotonically with the increasing laser intensity, and the velocity of the excited shock is roughly two times of the forward velocity of the laser front in a broad laser intensity regime ($4<a<10$).
It illuminates that in a near critical density plasma the collisionless shock generation is not sensitive to the laser intensity.

If the laser intensity $a>20$; however, we find no obvious evidence of the shock generation in our simulations. This is because an ultra-intense laser pulse will propagate too fast in a near critical density plasma due to the strong relativistic transparency \cite{WengNJP}, so there's no enough time for the electrons and ions to accumulate sufficiently ahead of the laser front. As a result, the laser pulse can propagate very fast during the whole laser-plasma interaction, and the ion reflections become very difficult in this case. On the other hand, if the initial laser intensity is too low, the forward velocity of the laser front will be not so high in comparison with the plasma thermal velocity. So it also becomes difficult to excite a fast shock in the background plasma if $a\leq3$ as shown in Fig. \ref{figvelocity}.

\subsection{Comparison between different density regimes}

\begin{figure}
\centering
\includegraphics[width=0.5\textwidth]{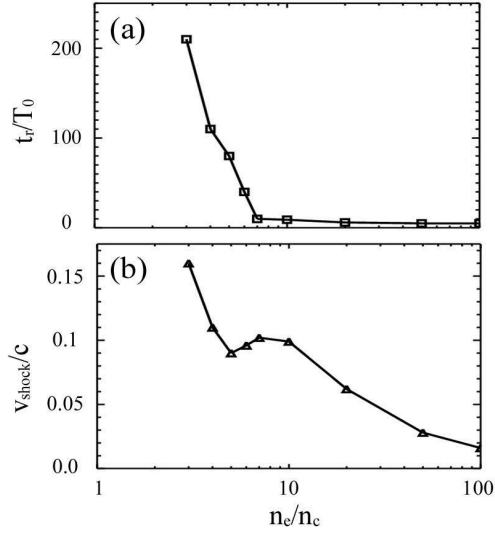}
\caption{(a) The response time $t_{r}$ for the shock formation and (b) the shock speed $v_{shock}$ as functions of the initial target density. The laser parameters are same to those in Fig. \ref{figRT2D}.} \label{figVsTs}
\end{figure}

It is important to notice that a considerable long response time is required before the laser pulse to excite the collisionless shock in the relativistically transparent plasmas. Fig. \ref{figVsTs}(a) shows that the response time for the shock formation decreases monotonously with the increasing initial target density. In the target with a density $n_e<7n_c$, the laser propagation is dominated by the relativistic transparency at the beginning. Therefore, at first there is only the accumulation of ions by the charge separation field, and the accumulation is too rarefied to excite the shock in the background plasma. With the accumulation of the ions ahead of the laser front, however, the laser pulse and the charge separation field will be slowed down and capable of accumulating more ions. Finally, the flow of accumulated ions will become dense enough to excite a collisionless shock in the background plasma. The lower the initial target density is, the longer response time it takes for the laser pulse to excite the collisionless shock. For targets with densities $n_e\leq 2n_c$, collisionless shock formation hasn't been observed in our PIC simulation up to $t=300T_0$.
In contrast, the collisionless shock can be quickly excited in a few laser wave cycles in the regime $n_e > 7n_c$ dominated by the HB. Taking into account the rising time of the laser pulse, the collisionless shock can be considered to be excited immediately in this regime.
Due to the transition of laser propagation mechanism from the relativistic transparency to the hole-boring,
we also notice that the shock speed increases slightly (but abnormally) with the increasing target density in the region $5n_c \leq n_e \leq 7n_c$ in Fig. \ref{figVsTs}(b). In the case of $n_e=5n_c$, the laser pulse at first can penetrate into the target via the relativistic transparency, but such a penetration is highly dissipative \cite{WengPoP}. Therefore, the laser pulse has been greatly weakened in the target. While the hole-boring dominates the laser pulse propagation nearly from the beginning in the case of $n_e=7n_c$, so the laser pulse can not penetrate deep into the target and becomes less dissipated. As a result, the speed of the excited shock in the case of $n_e=7n_c$ is faster than that in the case of $n_e=5n_c$.

\begin{figure}
\centering
 \includegraphics[width=0.5\textwidth]{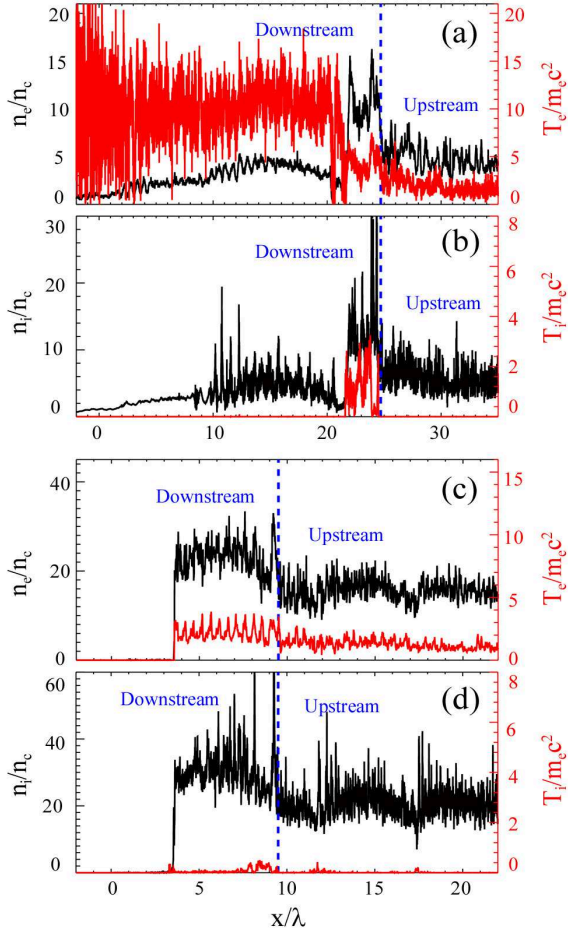}
\caption{Distributions of densities (black lines) and temperatures (red lines) of electrons and ions at $t=150T_0$ in the interaction of an intense laser pulse with (a)(b) a relativistic transparent target of $n_e=5n_c$ and (c)(d) a solid target of $n_e=20n_c$. The downstream and upstream regions of the shock structures are roughly separated by the blue dashed lines. The laser parameters are same to those in Fig. \ref{figRT2D}.} \label{figNeTe}
\end{figure}

We notice that the shock speed is as high as $v_{shock} \simeq 0.09c$ with the density $n_e=5n_c$.
This high shock speed may be attributed to the combination of the fast laser propagation and the highly heated plasma. The former provides a fast piston to excite the shock, while the latter results in a high ion acoustic speed.

As shown in Fig. \ref{figNeTe}(a), the plasma electron temperature $k_B T_{e0}$ in the upstream region is about $3m_ec^2$ for the relativistic transparency case $n_e=5n_c$. Compared with this high electron temperature, the ion temperature $k_B T_{i0}$ in the upstream region is negligible. Consequently, the ion acoustic speed can be estimated as $c_s\simeq\sqrt{k_{B}T_{e}/m_i} \simeq 0.040c$.
From the ion acoustic speed and the shock speed, we can estimate that the Mach number is about $M\simeq 2.25$.
In contrast, we find that the electron temperature in the upstream region is only about $1.2m_ec^2$ in a solid target case with $n_e=20n_c$. So the corresponding ion acoustic speed is only about $0.026c$. Considering the speed of the shock $v_{shock} \simeq 0.06c$ excited in this case, we can get the Mach number $M\simeq 2.31$. Although the Mach numbers of the shocks in the relativistically transparent and solid target cases are very close to each other, the shock speeds are considerably different due to the different ion acoustic speeds.

\subsection{Rankine-Hugoniot relations}

Substituting the Mach number $M=2.25$ into the jump conditions for an ideal shock \cite{Drake}, we can get the density and the pressure ratios between the downstream and the upstream regions as
\begin{eqnarray}
\frac{\rho_{1}}{\rho_{0}}= \frac{M^2(\gamma+1)}{M^2(\gamma-1)+2} \simeq 1.7, \label{densityRatio} \\
\frac{P_1}{P_0} =\frac{ 2\gamma M^2-(\gamma-1)}{\gamma+1} \simeq 7.1, \label{pressureRatio}
\end{eqnarray}
where the mass density $\rho=m_i n_i$, the pressure $P=n_ik_BT_i + n_e k_B T_e$, the subscript zero (one) denotes the upstream (downstream) variables, and $\gamma=1+2/f=3 $ is used with the degree of freedom $f=1$ for a 1D fully ionized plasma.
As shown in Figs. \ref{figNeTe}(a) and \ref{figNeTe}(b), the electron (ion) number densities in the upstream and downstream regions are about $n_{e0}\simeq5 n_c$ ($n_{i0}\simeq5 n_c$) and $n_{e1}\simeq10 n_c$ ($n_{i1}\simeq10 n_c$), respectively. And the electron (ion) temperatures in the upstream and downstream regions are about $k_B T_{e0}\simeq3 m_ec^2$ ($k_B T_{i0}\simeq0$) and $k_B T_{e1}\simeq 5 m_ec^2$ ($k_B T_{i1}\simeq 2 m_ec^2$), respectively.
From these simulation results, we can estimate the density ratio $\rho_{1}/\rho_{0}\simeq 2.0$ and the pressure ratio $P_1/P_0 \simeq 4.7$, which roughly agree with the prediction values given by Eqs. (\ref{densityRatio}) and Eqs. (\ref{pressureRatio}).

Furthermore, we find that the densities, pressures, and internal energies of the upstream and the downstream obtained from the simulation roughly satisfy the following Rankine-Hugoniot relations \cite{Eliezer}
\begin{eqnarray}
&&\rho_0 u_s = \rho_1 (u_s-u_p), \\
&&\rho_0 u_s u_p = P_1- P_0,  \\
&&\rho_0 u_s (E_1-E_0+u^2_p/2)=P_1 u_p, \label{RHrelations}
\end{eqnarray}
where the shock velocity $u_s\simeq 0.09c$, the piston velocity $u_p=v_{laser}\simeq0.05c$, and the internal energy per unit mass $E=k_B(n_iT_i + n_e T_e)/(m_in_i)$.

\subsection{Ion acceleration}
\begin{figure}
 \centering
 \includegraphics[width=0.5\textwidth]{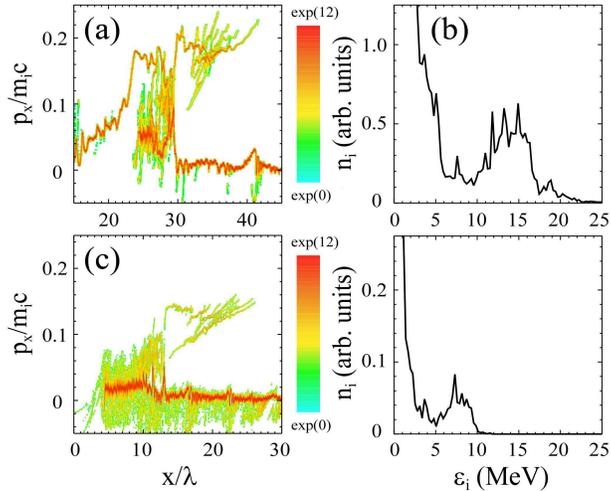}
\caption{(a) Ion distribution in the $p_x-x$ phase space and (b) ion energy spectrum at $t=200T_0$ in the interaction of an intense laser pulse with a relativistic transparent target;
and (c) and (d) are the corresponding ion distribution and energy spectrum in the case with a solid target.
The laser and target parameters are same to those in Fig. \ref{figNeTe}.
} \label{figPhase}
\end{figure}

As an important consequence of the high shock speed, the ions can be accelerated to very high energies in the laser interaction with a RT plasma.
As shown in Figs. \ref{figPhase}(a) and \ref{figPhase}(b), the center momentum and energy of shock-accelerated ions in the RT ($n_e=5n_c$) case are about $0.2 m_ic$ and $15$ MeV, respectively. While in the higher overdense target ($n_e=20n_c$) case they are only about $0.12 m_ic$ and $7.5$ MeV, respectively.
Furthermore, the yield of energetic ions can also be greatly enhanced in the RT case.
As indicated in Figs. \ref{figPhase}(b) and \ref{figPhase}(d), the number of energetic ions with $\varepsilon_i \geq 10$ MeV in the case of $n_e=5n_c$ is about 4 times larger than the number of energetic ions with $\varepsilon_i \geq 5$ MeV in the case of $n_e=20n_c$.

\section{\label{sec:level3}Conclusion and discussion}

Laser-driven collisionless shock formation and the consequent ion acceleration have been identified in the near critical density plasmas.
Comparing with a highly overdense target, an intense laser pulse can propagate much faster in a near critical density plasma, so does the laser-driven collisionless shock.
Consequently, a part of ions can be reflected and accelerated to very high energies.
However, in a near critical density plasma it may take a considerable long response time before the generation of the collisionless shock.
More important, we find that the longitudinal electric field related to the collisionless shock formation is strongly time-oscillating not only in the amplitude but also in the profile shape in this case.
With such a strongly time-oscillating electric field, the ions can be reflected even if the peak of the electrostatic potential is smaller than the ion kinetic energy in the shock frame.

In the above content, we use the dimensionless variables for the laser and target parameters.
For the employed laser intensity $a_0$=10 ($I\lambda^2 \simeq 1.37 \times 10^{20}$ W/cm$^2 \cdot \mu$m$^2$), it has already been achieved in the experiments using Nd:YAG or Ti:Sapphire laser pulses \cite{Zhang}. In order to excite a collisionless electrostatic shock in a near critical density plasma, however, the duration of the laser pulse should also be longer than a few hundred femtoseconds (about 200 hundred wave periods). Therefore, the relevant experiments might be conducted only in the picosecond petawatt laser facilities, such as LFEX \cite{PWLasers}.
Concerning the near critical density targets, they might be achieved with tunable density profiles by the advanced micro/nano-fabrication technologies \cite{WengSR}.
For instance, targets with the density range of 1-1000 mg/cm$^3$ can be fabricated by the pulsed laser deposition technique \cite{Zani}.
With the idea of the layer-by-layer nanoarchitectonics, nanotube films have been recently structured on the nanometer scale to provide near critical density targets \cite{Bin}. Alternatively, the near critical density targets might be also realized by the combination of highly compressed gas targets and intense infrared laser pulses \cite{Palmer,Tresca}.

The analyses in this paper are based on one-dimensional PIC simulations. Previous studies showed no qualitative difference between one- and multi-dimensional simulations of laser-driven collisionless shock formation in highly overdense targets \cite{He,Macchi,Silva}. However, in the near critical density regime it may take a longer response time before the shock formation in multi-dimensional simulations, since the laser transverse ponderomotive force will push particles out radially and slow down the accumulation of particles ahead the laser front. On the other hand, radial shocks might be excited by the laser transverse ponderomotive force too \cite{Wei}.

\section*{Acknowledgments}
\addcontentsline{toc}{section}{Acknowledgments}

This work was supported in part by the National Basic Research Program of China (Grant No. 2013CBA01504)
and the National Natural Science Foundation of China (Grant Nos. 11675108, 11405108, 11421064, 11129503, 11374210 and 11375262). SMW and MC appreciate the supports from National 1000 Youth Talent Project of China. ZMS acknowledges the support of a Leverhulme Trust Research Project Grant at the University of Strathclyde.
Simulations have been carried out on the PI cluster of Shanghai Jiao Tong University.


\section*{}

\end{document}